\documentclass[review]{elsarticle}

\usepackage{lineno,hyperref}
\usepackage[tight]{units}
\modulolinenumbers[5]

\def \be {\begin{equation}}
\def \ee {\end{equation}}

\journal{ArXiv}

\begin{document}

\begin{frontmatter}

\title{Boundary layer coupling of solid particles in water in an ultrasonic field}

\author{D. M. Forrester\fnref{myfootnote}}
\address{Chemical Engineering Department, Loughborough University, Loughborough, United Kingdom}

\ead{d.m.forrester@lboro.ac.uk}



\begin{abstract}
In an ultrasonic field the region around a solid particle experiences small, localised velocity, pressure, and thermal fluctuations that decay rapidly over short length scales. Herein, we investigate the overlap of the rapidly decaying fields between two silica particles in water in the case where the particles are aligned with each other in the direction of the applied field. We explore the velocity, pressure, temperature, and vorticity in the region of the particles. We discuss the coupled particle effects in ultrasonic waves as particles begin to agglomerate or become more concentrated. The analysis is conducted in the long wavelength regime for particles of diameter $\unit[500]{nm}$ and frequency $\unit[9.7]{MHz}$.  
\end{abstract}

\begin{keyword}
Coupled particles \sep Vorticity \sep Decay fields \sep Boundary layer effects
\end{keyword}

\end{frontmatter}


\section{Introduction}
Use of ultrasonic techniques are wide-spread throughout the industrial, medical, and research sectors \cite{Lifka2003,Payne1985,Poveybook,Microstructure2017}. In medicine, diagnostic imaging of the internal body structures is commonly undertaken using ultrasonography. Safety protocols are in place to make sure that sonographers do not use too much ultrasonic power when carrying out biological imaging (e.g. an ultrasound scan of a fetus) but little is actually known about possible cellular interactions with the applied field. For this reason localised field effects in confined particle spaces must be studied. Steps are commonly implemented, however, to avoid tissue damage  that may arise through overheating and mechanical stress, but fundamental changes to cellular processes can occur in response to even very small mechanical forces, as small as a trillionth of a newton \cite{ForresterPSSa2014}. With a continued drive towards using minute particles in vivo to deliver drugs \cite{Huebsch2014,Mullick2017}, stimulate cellular responses \cite{Mitragotri2005}, and to act as imaging modalities \cite{elastography2017}, Ultrasonic interactions with micro- and nanoparticles becomes increasingly important to understand. In most micro and nano confined structures novel interfacial effects are commonly found - e.g. localisation of light in subwavelength regions - surface plasmonics \cite{ForresterSCiRepWG2016}. Thus, we investigate the localised fields around a pair of silica particles in water. We show the overlap of these fields and discuss their implications for larger, more concentrated systems. The understanding of the visco-elastic boundary layer effects forms a basis for future studies of ultrasonic interactions with cells and to understand possible toxicity effects from environmental nanoparticles in plants and mammals.   

\section{Two coupled silica particles in water}
The localised scattered fields of two silica particles of $\unit[500]{nm}$ diameter are investigated with respect to the thermal and shear decay fields generated as a consequences of their material property contrasts with water. At the boundary between water and silica there are the two exponentially decaying fields, defined to have wavelengths \cite{Challis2005},
\be
\lambda_S=2\pi \sqrt{\frac{\mu}{\pi \rho f}},
\label{Eqn:SWL}
\ee 
and
\be
\lambda_T=2\pi \sqrt{\frac{\kappa}{\pi \rho f C_p}},
\label{Eqn:TWL}
\ee  
where $f$ is frequency, and $\lambda_S$ and $\lambda_T$ are the shear and thermal wavelengths, respectively. The physical material properties are listed in Table \ref{tbl:Table1}.    
\begin{table}[h]
\small
  \caption{Physical parameters of silica and water (at $\unit[25]{^o C}$)}
  \label{tbl:Table1}
\begin{center}
\begin{tabular}{ |c|c|c|c| } 
\hline
 & Silica & Water \\
\hline
Specific heat capacity, $C_p$ ($\unit[]{J/kg.K}$) & - & 4179 \\ 
Shear viscosity, $\mu$ ($\unit[]{Pa.s}$) & - & 0.000891 \\
Ratio of specific heats, $\gamma$ & - & 1.007 \\
Density, $\rho$ ($\unit[]{kg/m^3}$) & 2200 & 997 \\
Bulk modulus, $K$, ($\unit[]{GPa}$) & 37.16 & - \\
Shear modulus, $G$, ($\unit[]{GPa}$) & 30.9 & - \\
Thermal conductivity, $\kappa$, ($\unit[]{W/m.K}$) & - & 0.595 \\
Speed of sound, $c$, ($\unit[]{m/s}$) & 5968 & 1497 \\
Bulk viscosity, $\mu_B$, ($\unit[]{mPa.s}$) & - & 2.47 \\
Coeff. of thermal expansion, $\alpha$, ($\unit[]{1/K}$) & - & 0.00021 \\
\hline
\end{tabular}
\end{center}
\end{table}
The silica spheres are examined with $k_S r\approx 1-2$ (where $k_S$ is the shear wavenumber, and $r$ is the particle radius), which is the condition defined in previous works to identify the largest influences of shear fields in concentrated solid/liquid systems \cite{ForresterNS,ShearTheory2017}. We examine this particular condition because we wish to demonstrate the physics that occurs locally in such concentrated microfluids. Thus, we demonstrate the effects at $\unit[9.7]{MHz}$, which gives a value of $k_S r$ equal to $\unit[1.46]{}$. The particles are simulated using Comsol Multiphysics \cite{Comsol53} and the Thermoviscous Acoustics package (frequency domain) for the liquid phase coupled to the Solid Mechanics module for the solids. The details of these modules and their multiphysics combinations can be found in the Acoustic Module User's Guide (Ref. \cite{ComsolTVA}) and follow the Epstein-Carhart and Allegra-Hawley formulations (see Ref. \cite{EpsteinCarhart1953,AllegraHawley1972} for details). The model we use is 2D axis-symmetric in order to reduce the computational time. Even so, this simulation took nine days to complete. The particles sit in an off-set position with their centres at $z=13\times \lambda_S \pm 1.5r$. Thus, the separation distance between the edges of the particles is equal to the radius, $\unit[250]{nm}$. The thermoviscous domain is set around the particles in a region stretching 13 shear wavelengths from the central position in all directions (with an outer perfectly matched layer of thickness $6.5\lambda_S$ \cite{ComsolTVA}. A plane pressure wave with amplitude $\unit[0.1]{MPa}$ propagates
from the zero coordinate in the z direction. A triangular mesh is applied with maximum element size one fiftieth of a shear wavelength. The computer used to run the simulations has a CPU: Intel(R) Xeon(R) CPU E5-2630 v3 @ 2.40GHz, 16 cores, and 256 Gb of RAM. The number of degrees of freedom solved for was $18,220,504$.  

\section{Boundary effects leading to coupling}
In the vicinity of the solid particles there is the development of two forms of local field: those from visco-inertial oscillations and those from thermal pulsations of the scatterer \cite{Challis2005}.  
\begin{figure*}[ht!]
\centering
\includegraphics[width=12cm,keepaspectratio]{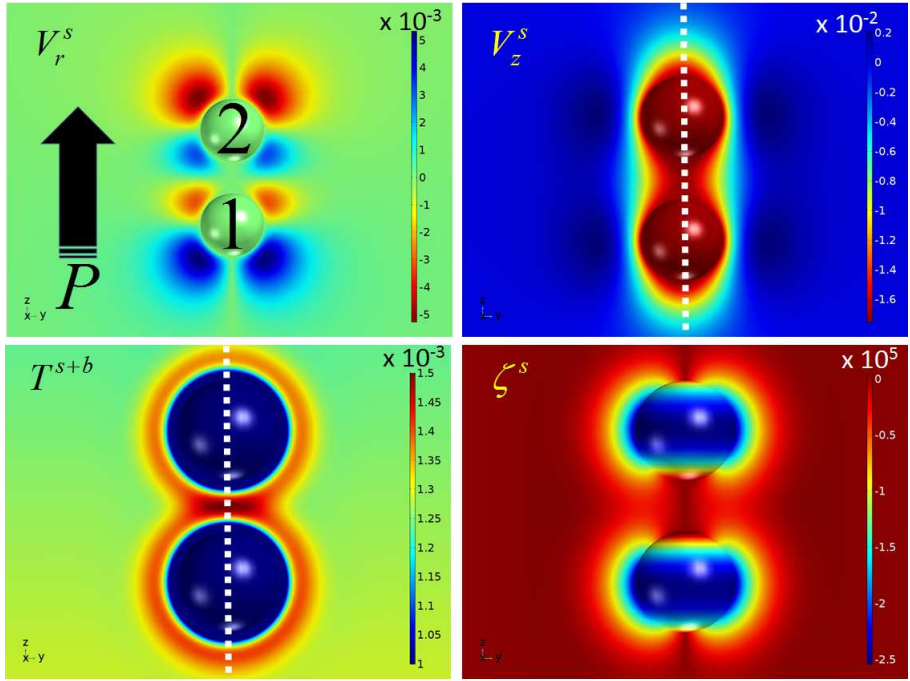}%
\caption{\label{Fig1} The scattered (s) fields around two aligned silica spheres of diameter $\unit[500]{nm}$ in an applied field with frequency $\unit[9.7]{MHz}$ and amplitude $\unit[0.1]{MPa}$. Top left: The $r$ component of the total acoustic velocity. Top right: The $z$ component of the total acoustic velocity. Bottom left: Monopolar thermal fields form (scattered plus background (b) shown) and overlap around each particle. Bottom right: The vorticity in the vicinity of the particles.}%
\end{figure*}
Figure \ref{Fig1} (top two plots) demonstrates the scattered velocity fields typically found around the particles. There is a quadrupolar appearance to the smaller $V_r$ component with polarised upper and lower lobes, at $45$ degree increments in the plane, and local velocities reaching $\pm\unit[5]{mm/s}$. The $V_z$ component dominates and takes a ``peanut shape'' around the particles due to the overlap of the fields belonging to each particle (velocities reach $\unit[20]{mm/s}$). The total temperature fluctuation is about a milli-kelvin around the particles, as shown in Fig. \ref{Fig1}, with an enhancement in the space directly between them (to about $\unit[1.5]{mK}$). The vorticity, given as a function of the velocity components ($\zeta=\frac{\partial V_{r,s}}{\partial z}-\frac{\partial V_{z,s}}{\partial x}$), shows coupling in the z-direction, but with the largest components in the radial direction very close to the particle surface.  The shear decay field has a wavelength of $\lambda_S=\unit[1.076]{\mu  m}$, whereas for the thermal decay field $\lambda_T=\unit[0.43]{\mu  m}$, as found through Eqn's (\ref{Eqn:SWL}-\ref{Eqn:TWL}) and verified through the FEM modelling.       
\begin{figure*}[ht!]
\centering
\includegraphics[width=12.5cm,keepaspectratio]{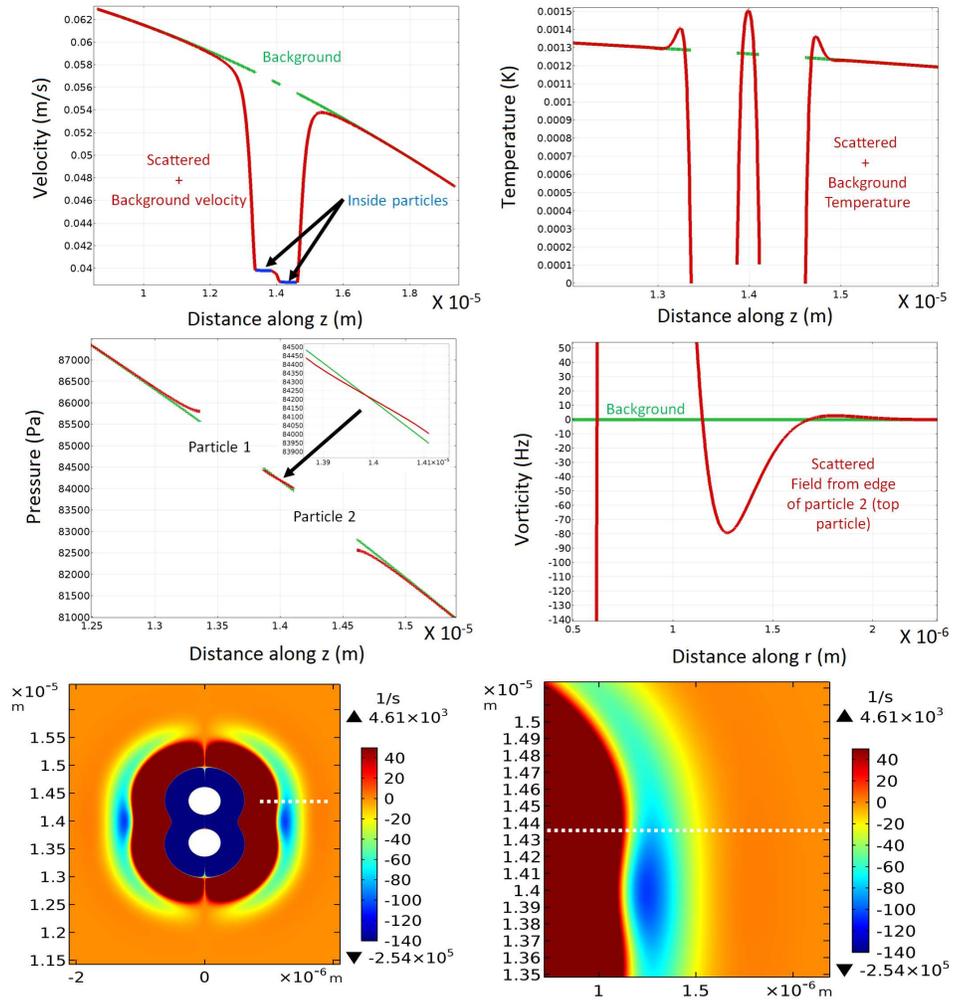}%
\caption{\label{Fig2} The scattered (red) and background fields (green) around the particles. Top left: The velocity $V_z$ along the z-axis that cuts through the centres of the particles. Top right: The temperature fluctuations around the particles due to the thermal field. Middle left: The pressure field fluctuations (red) along the z axis, superimposed over the background field pressure (green). Middle right: The oscillating azimuthal vorticity along the radial direction (path marked by the white line in the bottom left plot of vorticity)  of the second particle in the pressure field (top particle).  Bottom left: Vorticity field scaled to match that shown in the middle right plot. Bottom right: Magnification of the vorticity region in the bottom left image. The field propagates in the $+z$ direction.}%
\end{figure*}
The visualisation of the fields provided by Fig.\ref{Fig1} assists in interpretation of Fig. \ref{Fig2}. The white dashed lines in Fig. \ref{Fig1} also correspond to the directions indicated on the x-axes of Fig. \ref{Fig2} (top plots and middle left). Figure \ref{Fig2} (top left) demonstrates that with the inclusion of thermal and viscous effects the local velocities due to scattering are $\unit[31-34]{\%}$ lower than without (green line). The temperature fluctuations (Fig. \ref{Fig2}, top right)  exhibit a peak in the confined space between the particles, with the increase in temperature related to the separation distance and the thermal wavelength. The temperature changes at the particles boundaries due to the scattered field are $\approx \unit[1]{mK}$ and over several thermal wavelengths decay to the background level. The middle left plot in Fig. \ref{Fig2} illustrates the pressure amplitude along the z-direction. The difference in the pressures found with (red lines) and without (green lines) inclusion of scattering is as much as $\approx \pm \unit[250]{Pa}$ at the edges of the particles that are furthest from one another and $\approx \pm \unit[50]{Pa}$ in the confined space between them. Figure \ref{Fig2} highlights the vorticity in the radial direction around the particles due to the scattered shear fields. Without inclusion of shear waves, there is no vorticity.          

\section{Implications for more concentrated systems}
\begin{figure*}[ht!]
\label{Eqn:prad}
\centering
\includegraphics[width=12cm,keepaspectratio]{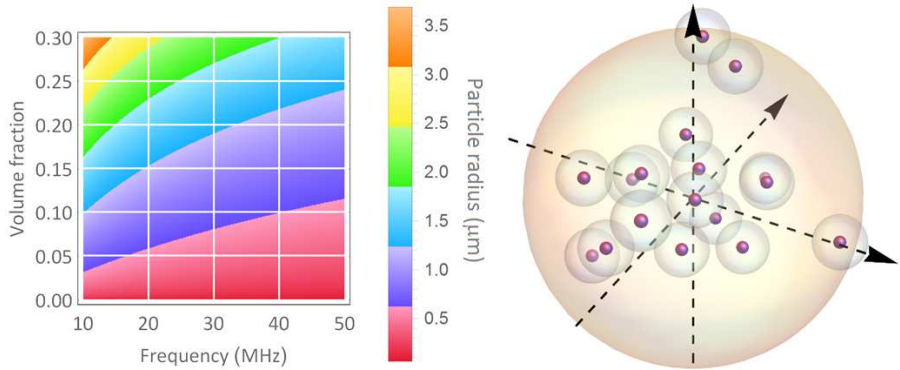}%
\caption{\label{Fig3} Particle sizes where viscous effects become important as a function of concentration (volume fractions, $0.001-0.3$ and ultrasonic frequency $\unit[10-50]{MHz}$ shown). The particle size is indicated by the colouration in the left hand plot and its colour bar at the side.   The plot is for particles in water at $\unit[25]{^o C}$. To the right is a depiction of $r=\unit[44]{nm}$ particles at $\unit[100]{MHz}$ with volume fractions of $0.001$ in a spherical volume of radius $3(\lambda_S+r)$. Around each particle is the region where shear effects become important.}%
\end{figure*}

This letter has highlighted two modes of coupling for solid particles in liquids when subjected to ultrasonic fields. The thermal and shear mode scattering is analysed using finite element modelling in order to better understand the localised field interactions due to shear and thermal effects. Indeed many traditional methods for modelling scattering from solid particles in fluids fail to account for the overlap of the thermal and visco-inertial boundary layers (e.g. \cite{LloydBerry1967}) and as the system becomes more concentrated there develop extra multiple scattering events and the traditional models break down. We can see from the simple two particle problem that the overlap of the thermal and viscous fields is of importance within one shear wavelength of the particle boundaries (e.g. see Fig. \ref{Fig2}, where vorticity has dropped to almost zero after a single shear wavelength). Indeed, setting the criterion for strong shear wave influences on the system to be $\psi\leq\lambda_S$, where $\psi$ is the average separation distance between the particles  (boundary to boundary), it is possible to estimate the particle sizes, concentrations, and frequencies where the overlap becomes important. We define the average separation between particles to be (similar to Ref. \cite{Chanamai1999} for emulsions) 
\be
\psi = rg,
\label{Eqn:sep}
\ee
where,
\be
g=\left[\left(\frac{\phi_{rcp}}{\phi}\right)^{1/3}-1\right],
\ee
$\phi$ is the volume fraction of spheres in water and $\phi_{rcp}$ is the random close packed volume fraction ($\approx 0.64$, for hard spheres). The relevance of the overlap effects increases when
\be
r\leq  \sqrt{\frac{4\pi \mu}{\rho f g^2}}
\ee
This approximation gives some insight into how coupled particle effects due to the decay fields emerge with different particle sizes. For example, in Fig. \ref{Fig3} particles with radii of $\unit[0.7]{\mu  m}$ will begin to be influenced by the overlap of the visco-inertial boundary layers at volume fractions of $0.04$ at $\unit[10]{MHz}$ ($\psi\leq\unit[1.06]{\mu  m}$) or $0.136$ at $\unit[50]{MHz}$ ($\psi\leq\unit[0.47]{\mu  m}$). A smaller nanoparticle has quite different values; e.g with a $r=\unit[44]{nm}$ particle at $\unit[100]{MHz}$, volume fractions of $0.001$ enable the nearest neighbours to be in the vicinity of one shear wavelength.          
Figure \ref{Fig3} shows the particle volume fractions at which concentrations of different sized particles (nano to microscale) at different frequencies begin to be couple through the overlap of the shear wave fields. For the particles of $\unit[500]{nm}$ diameter, as analysed using the FEM method, at a frequency of $\approx \unit[10]{MHz}$, the particle would experience interaction when separated by one $\lambda_S$ at volume fraction $\approx 0.0045$. We analysed the the two particles separated by $r$, which according to our approximation would be the average separation distance for a system of $\approx 13\%$ volume percent. In our previous experimental investigation of silica of various sizes  in water, including a nominal size of $\unit[500]{nm}$, using ultrasound spectroscopy, we determined that there was clear evidence for the overlap effect through conversion of the compressional wave to shear waves and back to the compressional wave at the particle/liquid boundaries. At $\approx 13-20\%$ volume percent and $\approx \unit[10]{MHz}$ (see Figure 3 of reference \cite{ForresterNS}) multiple scattering theory including shear effects matched the found attenuation spectra far better than without. Thus, the boundary layer fields are of high importance to those involved in sonography, particle characterisation, and medical analyses. It is worth noting, that the pressure wave amplitude can be varied in order to increase or decrease the thermal fields. Here, at an initial pressure amplitude of $\unit[0.1]{MPa}$ the change to the temperature is relatively small and inconsequential - making the shear mode dominant. The small thermal fluctuations of $mK$ size make the particles of this analysis compliant with FDA regulations banning increases of temperature higher than $1mK$ \cite{Mullick2017}. However, regularly higher amplitude pressures are applied in ultrasound guided drug delivery, for example, albeit at lower frequencies. Thus, future investigations should cover a larger span of particle sizes, pressure amplitudes, and frequencies using FEM. The current work is relevant to those involved in shallow biological imaging and diagnostics.    

\section{Conclusions}
Our numerical experiments using finite element modelling have revealed the nature of the ultrasonically induced coupling of solid particles, such as silica, in water. We examined the case of the two particles aligned in the ultrasound field. It is known that these effects have a large influence on the attenuation of the compressional wave in complex media. We also approximated the concentrations, frequencies and particle sizes where the boundary layer effects begin to emerge. We conclude that finite element modelling can be used to enable understanding of the pressures, forces, thermal and velocity fields in low to high concentration systems and will help develop the next generation of ultrasonic imagining and characterisation in ways currently impossible due to difficulties in experimentally observing the local fields.

\section{Acknowledgements}  

The author thanks the Engineering and Physical Sciences Research Council for funding under grant $EP/M026302/1$.


\section*{References}

\begin{thebibliography}{10}
\expandafter\ifx\csname url\endcsname\relax
  \def\url#1{\texttt{#1}}\fi
\expandafter\ifx\csname urlprefix\endcsname\relax\def\urlprefix{URL }\fi
\expandafter\ifx\csname href\endcsname\relax
  \def\href#1#2{#2} \def\path#1{#1}\fi

\bibitem{Lifka2003}
J.~Lifka, B.~Ondruschka, J.~Hofmann, The use of ultrasound for the degradation
  of pollutants in water: Aquasonolysis – a review, Engineering in Life
  Sciences 3~(6) (2003) 253--262.
\newblock \href {http://dx.doi.org/10.1002/elsc.200390040}
  {\path{doi:10.1002/elsc.200390040}}.

\bibitem{Payne1985}
P.~A. Payne, Medical and industrial applications of high resolution ultrasound,
  J. Phys. E: Sci. Instrum. 18~(6) (1985) 465.
\newblock \href {http://dx.doi.org/10.1088/0022-3735/18/6/001}
  {\path{doi:10.1088/0022-3735/18/6/001}}.

\bibitem{Poveybook}
M.~J.~W. Povey, Ultrasonic Techniques for Fluids Characterization, 1st Edition,
  Academic Press LTd, California, USA, 1997.

\bibitem{Microstructure2017}
K.~Rajewska, D.~Mierzwa, Influence of ultrasound on the microstructure of plant
  tissue, Innovative Food Science \& Emerging Technologies 43 (2017) 117--129.
\newblock \href {http://dx.doi.org/10.1016/j.ifset.2017.07.034}
  {\path{doi:10.1016/j.ifset.2017.07.034}}.

\bibitem{ForresterPSSa2014}
M.~Forrester, F.~Kusmartsev, The nano-mechanics and magnetic properties of high
  moment synthetic antiferromagnetic particles, Physica Status Solidi a 211~(4)
  (2014) 884--889.
\newblock \href {http://dx.doi.org/10.1002/pssa.201330122}
  {\path{doi:10.1002/pssa.201330122}}.

\bibitem{Huebsch2014}
N.~Huebsch, C.~J. Kearney, X.~Zhao, J.~Kim, C.~Cezar, Z.~Suo, D.~J. Mooney,
  Ultrasound-triggered disruption and self-healing of reversibly cross-linked
  hydrogels for drug delivery and enhanced chemotherapy, Proc. Natl. Acad. Sci.
  U S A 111~(27) (2014) 9762--9767.
\newblock \href {http://dx.doi.org/10.1073/pnas.1405469111}
  {\path{doi:10.1073/pnas.1405469111}}.

\bibitem{Mullick2017}
S.~{Mullick Chowdhury}, T.~L. J.~K. Willmann, Ultrasound-guided drug delivery
  in cancer, Ultrasonography 36~(3) (2017) 171--184.
\newblock \href {http://dx.doi.org/10.14366/usg.17021}
  {\path{doi:10.14366/usg.17021}}.

\bibitem{Mitragotri2005}
S.~Mitragotri, Healing sound: the use of ultrasound in drug delivery and other
  therapeutic applications, Nature Reviews Drug Discovery 4 (2005) 255--260.
\newblock \href {http://dx.doi.org/10.1038/nrd1662}
  {\path{doi:10.1038/nrd1662}}.

\bibitem{elastography2017}
B.-G. Zhou, D.~Wang, W.-W. Ren, X.-L. Li, Y.-P. He, B.-J. Liu, Q.~Wang, S.-G.
  Chen, A.~Alizad, H.-X. Xu, Value of shear wave arrival time contour display
  in shear wave elastography for breast masses diagnosis, Scientific Reports 7
  (2017) 7036.
\newblock \href {http://dx.doi.org/10.1038/s41598-017-07389-0}
  {\path{doi:10.1038/s41598-017-07389-0}}.

\bibitem{ForresterSCiRepWG2016}
D.~M. Forrester, F.~V. Kusmartsev, Whispering galleries and the control of
  artificial atoms, Scientific Reports 6 (2016) 25084.
\newblock \href {http://dx.doi.org/10.1038/srep25084}
  {\path{doi:10.1038/srep25084}}.

\bibitem{Challis2005}
R.~E. Challis, M.~J.~W. Povey, M.~L. Mather, A.~K. Holmes, Ultrasound
  techniques for characterizing colloidal dispersions, Rep. Prog. Phys. 68~(7)
  (2005) 1541--1637.
\newblock \href {http://dx.doi.org/10.1088/0034-4885/68/7/R01}
  {\path{doi:10.1088/0034-4885/68/7/R01}}.

\bibitem{ForresterNS}
D.~M. Forrester, J.~Huang, V.~J. Pinfield, F.~Lupp\'e, Experimental
  verification of nanofluid shear-wave reconversion in ultrasonic fields,
  Nanoscale 8 (2016) 5497--5506.
\newblock \href {http://dx.doi.org/10.1039/C5NR07396K}
  {\path{doi:10.1039/C5NR07396K}}.

\bibitem{ShearTheory2017}
V.~J. Pinfield, D.~M. Forrester, Multiple scattering in random dispersions of
  spherical scatterers: Effects of shear-acoustic interactions, {J}. {A}coust.
  {S}oc. {A}m. 141 (2017) 649--660.
\newblock \href {http://dx.doi.org/10.1121/1.4974142}
  {\path{doi:10.1121/1.4974142}}.

\bibitem{Comsol53}
{COMSOL AB}, {COMSOL Multiphysics v. 5.3}, Stockholm, Sweden.

\bibitem{ComsolTVA}
{COMSOL AB}, {Acoustic Module User's Guide,Thermoviscous Acoustics Branch },
  Stockholm, Sweden, 2017.

\bibitem{EpsteinCarhart1953}
P.~S. Epstein, R.~R. Carhart, {T}he absorption of sound in suspensions and
  emulsions. i. water fog in air, {J}. {A}coust. {S}oc. {A}m. 25~(3) (1953)
  553--565.
\newblock \href {http://dx.doi.org/10.1121/1.1907107}
  {\path{doi:10.1121/1.1907107}}.

\bibitem{AllegraHawley1972}
J.~R. Allegra, S.~A. Hawley, {A}ttenuation of sound in suspensions and
  emulsions: Theory and experiments, {J}. {A}coust. {S}oc. {A}m. 51 (1972)
  1545--1564.
\newblock \href {http://dx.doi.org/10.1121/1.1912999}
  {\path{doi:10.1121/1.1912999}}.

\bibitem{LloydBerry1967}
P.~Lloyd, M.~V. Berry, {W}ave propagation through an assembly of spheres: Iv.
  relations between different multiple scattering theories, {P}roc. {P}hys.
  {S}oc. 91 (1967) 678--688.
\newblock \href {http://dx.doi.org/10.1088/0370-1328/91/3/321}
  {\path{doi:10.1088/0370-1328/91/3/321}}.

\bibitem{Chanamai1999}
R.~Chanamai, N.~Herrmann, D.~J. McClements, Influence of thermal overlap
  effects on the ultrasonic attenuation spectra of polydisperse oil-in-water
  emulsions, Langmuir 15~(10) (1999) 3418--3423.
\newblock \href {http://dx.doi.org/10.1021/la981195f}
  {\path{doi:10.1021/la981195f}}.

\end{thebibliography}
\end{document}